\documentclass[aps, prl, twocolumn, letterpaper,superscriptaddress]{revtex4-1}
\pdfoutput=1

\usepackage{amsmath}	
\usepackage{amsthm}		
\usepackage{amssymb}	
\usepackage{eufrak}
\usepackage{datetime}
\usepackage{graphicx}
\usepackage{color}
\usepackage{verbatim}
\usepackage{bm}
\usepackage{braket}
\usepackage{cancel}
\usepackage{xcolor}
\usepackage{slashed}
\usepackage{braket}
\usepackage{caption}
\usepackage{subcaption}
\usepackage{amsfonts} 
\usepackage{float}
\usepackage{dsfont}
\smallskipamount
\usepackage{feynmp}
\usepackage{letltxmacro} 
\usepackage{microtype}

\usepackage[colorlinks=true, citecolor=midblue, linkcolor=midblue, urlcolor=midblue]{hyperref}

\usepackage{setspace}

\usepackage{tikz,xcolor}
\definecolor{lime}{HTML}{A6CE39}
\DeclareRobustCommand{\orcidicon}{
	\begin{tikzpicture}
	\draw[lime, fill=lime] (0,0) 
	circle [radius=0.16] 
	node[white] {{\fontfamily{qag}\selectfont \tiny ID}};
	\draw[white, fill=white] (-0.0625,0.095) 
	circle [radius=0.007];
	\end{tikzpicture}
	\hspace{-2mm}
}
\foreach \x in {A, ..., Z}{
	\expandafter\xdef\csname orcid\x\endcsname{\noexpand\href{https://orcid.org/\csname orcidauthor\x\endcsname}{\noexpand\orcidicon}}
}

\settimeformat{ampmtime}

\definecolor{grey}{rgb}{0.4,0.4,0.4}
\definecolor{dullmagenta}{rgb}{0.4,0,0.4}
\definecolor{darkblue}{rgb}{0,0,0.4}
\definecolor{midblue}{rgb}{0,0,0.5}
\definecolor{midred}{rgb}{0.5,0,0}
\definecolor{orange}{rgb}{1,0.5,0}
\definecolor{lightbrown}{rgb}{0.75,0.5,0.25}
\definecolor{tan}{cmyk}{0.14,0.42,0.56,0}
\definecolor{djunglegreen}{cmyk}{0.99,0,0.52,0}
\definecolor{lightgreen}{rgb}{0,1,0}
\definecolor{olivegreen}{cmyk}{0.64,0,0.95,0.40}
\definecolor{midgreen}{rgb}{0.0,0.675,0.0}
\definecolor{darkgreen}{rgb}{0,0.5,0}

\newcommand{\vs}{\vspace}

\renewcommand{\.}{\hspace{0.5mm}}

\newcommand{\drm}{\ensuremath{\mathrm{d}}}

\newcommand{\Lcal}{\ensuremath{\mathcal{L}}}

\newcommand{\Ocal}{\ensuremath{\mathcal{O}}}

\renewcommand{\d}{\ensuremath{\mathrm{d}}}

\def\Msun{M_\odot}

\settimeformat{ampmtime}

\usepackage{float}

\makeatletter
\let\oldr@@t\r@@t
\def\r@@t#1#2{%
\setbox0=\hbox{$\oldr@@t#1{#2\,}$}\dimen0=\ht0
\advance\dimen0-0.2\ht0
\setbox2=\hbox{\vrule height\ht0 depth -\dimen0}%
{\box0\lower0.4pt\box2}}
\LetLtxMacro{\oldsqrt}{\sqrt}
\renewcommand*{\sqrt}[2][\ ]{\oldsqrt[#1]{#2}}
\makeatother

\newcommand{\FirstAffiliation}{\affiliation{
	Arnold Sommerfeld Center,
	Ludwig-Maximilians-Universit{\"a}t,
	Theresienstra{\ss}e 37,
	80333 M{\"u}nchen,
	Germany}}
 
\newcommand{\SecondAffiliation}{\affiliation{
	Max-Planck-Institut f{\"u}r Physik,
	F{\"o}hringer Ring 6,
	80805 M{\"u}nchen,
	Germany}}

\newcommand{\ThirdAffiliation}{\affiliation{
    Tsung-Dao Lee Institute and School of Physics and Astronomy,
    Shanghai Jiao Tong University, Shengrong Road 520, 201210
    Shanghai, China}}

\begin{document}

\title{Vortex Effects in Merging Black Holes and Saturons}

\author{Gia Dvali}
\FirstAffiliation
\SecondAffiliation

\author{Oleg Kaikov\!\orcidB{}}
\email{kaikov@mppmu.mpg.de}
\FirstAffiliation
\SecondAffiliation

\author{Florian K{\"u}hnel\!\orcidD{}}
\FirstAffiliation
\SecondAffiliation

\author{Juan Sebasti{\'a}n Valbuena-Berm{\'u}dez\!\orcidA{}} 
\email{juanv@mpp.mpg.de}
\FirstAffiliation
\SecondAffiliation

\author{Michael Zantedeschi\!\orcidC{}}
\email{zantedeschim@sjtu.edu.cn}
\ThirdAffiliation

\date{\formatdate{\day}{\month}{\year}, \currenttime}

\begin{abstract}
\noindent
Vorticity has recently been suggested to be a property of highly-spinning black holes. The connection between vorticity and limiting spin represents a universal feature shared by objects of maximal microstate entropy, so-called saturons. Using $Q$-ball-like saturons as a laboratory for black holes, we study the collision of two such objects and find that vorticity can have a large impact on the emitted radiation as well as on the charge and angular momentum of the final configuration. As black holes belong to the class of saturons, we expect that the formation of vortices can cause similar effects in black hole mergers, leading to macroscopic deviations in gravitational radiation. This could leave unique signatures detectable with upcoming gravitational-wave searches, which can thereby serve as a portal to macroscopic quantum effects in black holes.
\end{abstract}

\maketitle

\paragraph{\textbf{Introduction.}} Recently, it has been proposed that black holes may admit vortex structures~\cite{Dvali:2021ofp}. This proposal is supported by two separate lines of reasoning. First, such a possibility is rather natural within the description of a black hole as a condensate of soft gravitons at a quantum critical point~\cite{Dvali:2011aa, Dvali:2012en}, as in general, the Bose-Einstein condensates are known to exhibit the vortex structure. 

Secondly, the possibility of vortices within black holes is supported by an alternative reasoning that is independent of a particular microscopic proposal. Instead, this argument relies on the universal properties of the phenomenon of saturation. It has recently been argued~\cite{Dvali:2020wqi} that key black hole properties, such as the area-law of the entropy, a near-thermal evaporation, and a long time-scale of information retrieval, are not specific to black holes. Rather, they represent universal features of objects exhibiting the maximal microstate entropy permitted by a given quantum-field-theoretical (QFT) description. Such objects were referred to as {\it saturons}. Several explicit examples have been studied in a series of papers~\cite{Dvali:2019jjw, Dvali:2019ulr, Dvali:2020wqi, Dvali:2021rlf, Dvali:2021tez}, which fully confirm the above universality.

This remarkable correspondence between the black holes and other saturons gives us a double advantage. First, non-gravitational saturons emerge as interesting creatures in their own right, which can have spectacular consequences both for fundamental physics as well as for observations. In particular, they can have various cosmological implications~\cite{Dvali:2023xfz}.

At the same time, saturons can serve as a laboratory for understanding the existing features of black holes and predicting new ones, solely using the power of universality of saturation, without the need to enter into the technicalities of quantum gravity. 
\newpage

The present paper is devoted to one of such features: vorticity. The key starting point of our study is the following. First, it has been noticed in Ref.~\cite{Dvali:2021ofp} that the saturon spin is linked with its internal vorticity. Crucially, the vorticity sets the relation between the maximal spin $J^{\rm max}$ of a saturon and its entropy $S$,
\begin{equation}
\label{eq:maximalspinsat}
     J^{\rm max} = S\..
\end{equation}
Remarkably, this relation copies the analogous well-known relation for extremal black holes, $J_{\rm BH}^{\rm max} = S_{\rm BH}$.

In the case of a black hole, there exists no commonly accepted microscopic explanation of the above relation. However, in light of the black hole/saturon correspondence, it was suggested in Ref.~\cite{Dvali:2021ofp} that the underlying mechanism limiting the black hole spin must be vorticity, as is the case for saturons. This implies that vorticity is expected to be exhibited by highly-spinning black holes. 

Beyond the explanation of the upper bound on the black holes' spin, this conjecture also sheds light on other seemingly mysterious features of black holes with large spins. For example, it provides a rationale for the absence of Hawking emission in an extremal black hole: on the analog side of an extremal saturon, the absence of emission is due to the topological stability of the extremal vortex~\cite{Dvali:2021ofp}. 

Needless to say, the existence of vorticity in black holes is expected to have potentially observable consequences. In particular, vorticity could localize a magnetosphere around a black hole without the need for a specific accretion disk required in the standard scenario~\cite{McKinney:2012vh}. Correspondingly, powerful jets could be emitted~\cite{Blandford:1977ds}, providing a smoking gun for the vorticity.

In this letter, we study another important consequence of vorticity in black holes. Due to the macroscopic difference in the substructure of a black hole with or without a vortex, it is natural to expect a discontinuity in the gravitational radiation emitted by black hole mergers. 

In order to verify this idea, we study how the formation of vorticity in the final state affects the collision dynamics of saturons. As we will show, the presence of vorticity indeed leads to macroscopic deviations in the emitted radiation. In particular, this is due to a very high sensitivity of the energy output with respect to vortex formation.  Motivated by the universality of saturated objects, we propose that similar dynamics could take place in certain black hole mergers, strongly affecting the emitted gravitational radiation. Specifically, we expect this phenomenon to manifest itself in the form of delayed or suppressed gravitational radiation in case of vortex formation. 

In our analysis, we use the model of saturons in the form of vacuum bubbles, originally introduced in Ref.~\cite{Dvali:2020wqi}. These objects spontaneously break a global SU$(N)$ symmetry and are stabilized by the corresponding Goldstone charge~\cite{Dvali:2021tez}. In this sense, they represent a form of $Q$-balls~\cite{Friedberg:1976me, Coleman:1985ki}. Correspondingly, our results are independently motivated by understanding the cosmological and astrophysical consequences of $Q$-balls and other types of saturated solitons. 

We first proceed with the introduction of technical details of the model and present our findings regarding the collision process as well as vortex formation. Afterwards, we elaborate on the consequences of our results for black hole mergers. A summary of the present work, as well as videos of the simulated dynamics, can be found at these two~\href{https://www.youtube.com/watch?v=t29WUvZM-io}{URL1} and \href{https://youtu.be/zorkQSYCliU}{URL2}. Throughout this work we use units in which $c = \hbar = 1$.

\paragraph{\textbf{A Model for Saturons.}} 

In our analysis, we use a specific model of saturons within a renormalizable field theory, as introduced in Ref.~\cite{Dvali:2020wqi}. Stable saturon constructions without and with spin are given in Refs.~\cite{Dvali:2021tez} and~\cite{Dvali:2021ofp}, respectively.

The theory consists of a scalar field $\Phi$ in the adjoint representation of SU$( N )$ with the Lagrangian
\begin{equation}
\label{eq:lagphi}
    \Lcal[\Phi] = \frac{1}{2}\,{\rm{Tr}}\,(\partial_{\mu} \Phi)(\partial^{\mu} \Phi) - V( \Phi )\, .
\end{equation}
The potential $V( \Phi )$ is given by
\begin{equation}
\label{eq:potential}
    V( \Phi ) = \frac{\alpha}{2}\,{\rm{Tr}}\left[f\Phi -\Phi^{2} + \frac{\mathds{1}}{N}\,{\rm{Tr}}\,\Phi^{2}\right]^{2}\, ,
\end{equation}
with $\mathds{1}$ being the unit matrix, $\alpha$ a dimensionless coupling constant and $f$ the scale. Notice that the validity of QFT description in terms of $\Phi$, imposes the constraint~\cite{Dvali:2020wqi}
\begin{equation}
    \alpha \lesssim 1/N\,.
\end{equation}

Inspired by the analogy with black holes we shall work at large $N$ keeping in mind the double-scaling limit
\begin{equation}
\label{eq:doublesc}
    N\rightarrow\infty\, ,\quad \alpha\rightarrow 0\, , \quad \alpha\, N\rightarrow \Ocal(1)\, ,
\end{equation}
where the last expression determines the strength of the collective coupling. The model~\eqref{eq:lagphi} has multiple degenerate vacua satisfying the condition
\begin{equation}
\label{eq:vacua_condition}
    f\.\Phi^{b}_{\phantom{b}a} -
    \left(\Phi^{2}\right)^{b}_{\phantom{b}a} + \.\delta^{b}_{\phantom{b}a}\,\frac{1}{N}\,{\rm{Tr}}\.\Phi^{2} = 0\, .
\end{equation}
The different constant solutions to Eq.~\eqref{eq:vacua_condition} realize distinct patterns of symmetry breaking. 

We are interested in vacuum-bubble configurations that interpolate between the SU$(N)$ symmetric vacuum in the exterior and SU$(N - 1) \times $U$(1)$ vacua in the interior of the bubble. In the former vacuum, realized asymptotically, the theory is in the gapped phase, with mass-squared $m^{2} = \alpha \,f^{2}$. In the latter vacuum, localized in the interior of the bubble, the symmetry is spontaneously broken. Therefore, $2(N-1)$ Goldstone species exist in that region. A wall of tension $m^{3} / 6\mspace{2mu}\alpha$ and of thickness $1/m$ separates these two regions.

We build such configurations via the ansatz~\cite{Dvali:2021tez}
\begin{equation}
	\label{eq:rotation}
    \Phi = U^{\dagger} \phi\mspace{3mu} U\, ,
\end{equation}
in which 
\begin{equation}
\label{eq:phi} 
    \phi = \dfrac{ \rho( x ) }{ N }\.\text{diag}\big[( N - 1 ), - 1, \dotsc, - 1\big]\, ,
\end{equation}
and 
\begin{equation}
\label{eq:Umatrix}
	U = \exp\mspace{-3mu}\Big[ i\.\theta( x )\.\hat{T}/\sqrt{2}\mspace{1.5mu} \Big]\, .
\end{equation}
Above, $\hat{T}$ corresponds to generators broken by the ansatz~\eqref{eq:phi}, and $\theta(x)$ to the Goldstone modes. This leads to the effective leading-order (large-$N$ and up to second order in $\theta$) Lagrangian
\begin{equation}
	\label{eq:Lag}
	\Lcal = \frac{ 1 }{ 2 }\.\Big[\partial_{\mu}\rho\.\partial^{\mu}\rho
    + \frac{1}{2}\rho^{2}\.\partial_{\mu}\theta\.\partial^{\mu}\theta
    -\alpha\.\rho^{2}\mspace{1.5mu}(\rho-f)^{2}\Big]\..
\end{equation}
Here and below, the $N$-dependent factors are absorbed into the respective redefinition of the parameters. For simplicity, we consider the case in which a single broken generator is macroscopically occupied. In this effective theory, we are interested in two solutions. 

The first solution corresponds to the choice
\begin{equation}
\label{eq:bubblespherical}
    \rho(x) = \rho(r)\,,\quad \theta(x) = \omega\, t\,,
\end{equation}
describing a spherical bubble configuration (see Ref.~\cite{Dvali:2021tez}). The profile $\rho(r)$ obeys the asymptotic conditions $\rho( 0 ) \simeq f$ and $\rho( \infty ) = 0$. This configuration is stable thanks to the occupied Goldstone mode number{\,---\,}or charge{\,---\,}$Q$. According to the Noether theorem, this charge can be estimated as
\begin{equation}
\begin{split}
\label{eq:Qcharge}
   Q
	& =  i\;{\rm Tr}\left[ \int \d^{3}x\;
                \big[
                    \partial_{t} \Phi,\.\Phi
                \big]\,
                \hat{T}\right]\\[2mm]
    & = 2\pi\mspace{2mu}\omega\int \drm r\;r^{2}\.\rho^{2}( r )\simeq  \dfrac{ 2\pi }{ 3\alpha }\.m^{2}\.\omega\mspace{1mu}R^{3}
					\, ,
\end{split}
\end{equation}
where the second equality follows from ansatz~\eqref{eq:bubblespherical}. The mechanism of classical stability of the bubble shares similarities with $Q$-balls.

The distinctive feature is that the bubble exhibits an exponentially large microstate degeneracy due to $\sim N$ species of Goldstone modes localized within its interior (see Ref.~\cite{Dvali:2021tez} for more details). It was shown that in the region of parameter space
\vs{-1mm}
\begin{equation}
    \label{eq:saturationbubblecondition}
    \omega\sim m\sim 1/R\,,\quad N\sim Q\sim 1/\alpha\, ,
\end{equation}
the bubble saturates the upper bound on entropy imposed by the validity of QFT~\cite{Dvali:2021tez},
\begin{equation}
    S \sim \frac{1}{\alpha}\sim (R\,f)^{2} \sim M^{2}/f^{2}\, ,
\end{equation}
where $M$ is the energy of the configuration. In other words, in this regime, the bubble represents a saturon with features analogous to a black hole. In particular, the microstate entropy of the saturon bubble mimics the Bekenstein-Hawking entropy~\cite{Bekenstein:1973ur, Hawking:1975vcx} of a black hole, with $f$ replacing the role of the Planck mass.

The second solution corresponds to the case in which the bubble is pierced by a global vortex line~\cite{Vilenkin:2000jqa}. The construction is very similar to the one introduced for Abelian $Q$-balls considered in Ref.~\cite{Kim:1992mm} in $2 + 1$ dimensions and generalized to $3 + 1$ in Ref.~\cite{Volkov:2002aj}. In our case, we consider the axially-symmetric ansatz 
\begin{equation}
\label{eq:bubbleaxial}
    \rho = \rho (r,\.\chi)\,,\quad \theta = \omega\,t + n\,\varphi\, ,
\end{equation}
where $\chi,\varphi$ denote the azimuthal and the polar angles, respectively. Asymptotically, we have: $\rho(0,\.\chi) = 0 = \rho(\infty,\.\chi) = \rho(r,\.0) = \rho(r,\.\pi) = 0$. As in the previous case of a non-spinning solution, the bubble is stabilized by the Goldstone modes of frequency $\omega$. However, the same Goldstone modes now form a topological current: the Goldstone field exhibits an integer winding number $n$. This gives the spin of the saturon, 
\begin{equation}
\label{eq:Jt0phi}
    J = \int\d^{3}x\; T_{0\varphi} = n\, Q = n\,S\, ,
\end{equation}
where $T_{\mu \nu}$ is the energy momentum tensor of the configuration. The second equality follows from the ansatz~\eqref{eq:bubbleaxial}, and the third one from the saturation condition~\eqref{eq:saturationbubblecondition}.

It can be shown that in order to maintain the entropy saturation of the configuration, $n$ cannot be much greater than one. From this, the maximal-spin condition for rotating $Q$-ball-type saturons~\eqref{eq:maximalspinsat} is recovered~\cite{Dvali:2021ofp}. 

The explicit stationary solutions that follow from the Ans{\"a}tze~\eqref{eq:bubblespherical} and~\eqref{eq:bubbleaxial} can be found in Ref.~\cite{Dvali:2021tez}, and, Ref.~\cite{Zantedeschi:2022czs}, respectively, for the three-dimensional case. 
\vs{3mm}

\paragraph{\textbf{Vorticity in Saturon Mergers.}} We now proceed with the study of the collision of spherical bubbles [c.f.~Eq.~\eqref{eq:bubblespherical}] around the saturation point~\eqref{eq:saturationbubblecondition}. For practical numerical purposes, we set $N = 4$.

We boost the bubbles towards each other with different velocities and impact parameters. We observe that a sufficient condition for vortex formation is related to the relative phase between the two bubbles. Namely, in order to create a vortex, the chosen $\Delta \theta$ [see Eq.~\eqref{eq:bubblespherical}] should be equal to $\pi$. The reason is the following: The vortex ansatz for $\theta( x )$~\eqref{eq:bubbleaxial} at a fixed time gives opposite phase upon a rotation by $\pi$ in $\varphi$ for $n = 1$. Therefore, choosing the above offset for $\Delta\theta$ ensures proximity to the vortex bound state. Moreover, exactly at $\Delta \theta = \pi$ the merged bubble is subject to a ``point'' symmetry around its center, enforcing the presence of a vortex. As we shall see below, small deviations from the above condition only lead to temporal vortex formation.

Once the relative phase is chosen, the initial conditions determine the kinematics. For a broad range of initial velocities, between $0.2$ and $0.8$, there exists a wide range of impact parameters comparable to the bubble radius, ensuring vortex formation. For example, for bubble relative velocities between $0.3$ and $0.5$, an impact parameter between $0.3$ and $0.8$ times the bubble radius is sufficient. 
This shows no significant tuning is needed for the impact parameter.

The impact parameter, however, needs significantly more tuning for higher velocity. Moreover, the relevant range of parameters is affected by the thickness of the bubble wall as it seems generally easier to attain vorticity in the thin-wall regime. We believe this is happening due to the fact that for larger cores it is easier to generate high angular momentum at milder velocities, effectively resulting in a larger dynamical window. If the velocities are too high and/or the impact parameter is too large, the bubbles simply scatter off from each other.

For definiteness, we will focus on three cases that exemplify the relevance of our findings when extrapolated to black holes~\footnote{We postpone a detailed analysis and characterization of the scattering dynamics for future work~\cite{Vortex-Future-Paper}.}. These cases correspond to different choices of $\Delta \theta = 0,\,0.95\,\pi,\,\pi$, while keeping all other parameters unchanged. We will refer to the first, second, and third cases as the {\it no-vortex}, {\it ejected-vortex} and {\it vortex} regimes, respectively. The reason behind these names will become apparent below.

\begin{figure*}
    \includegraphics[width=.85\textwidth]{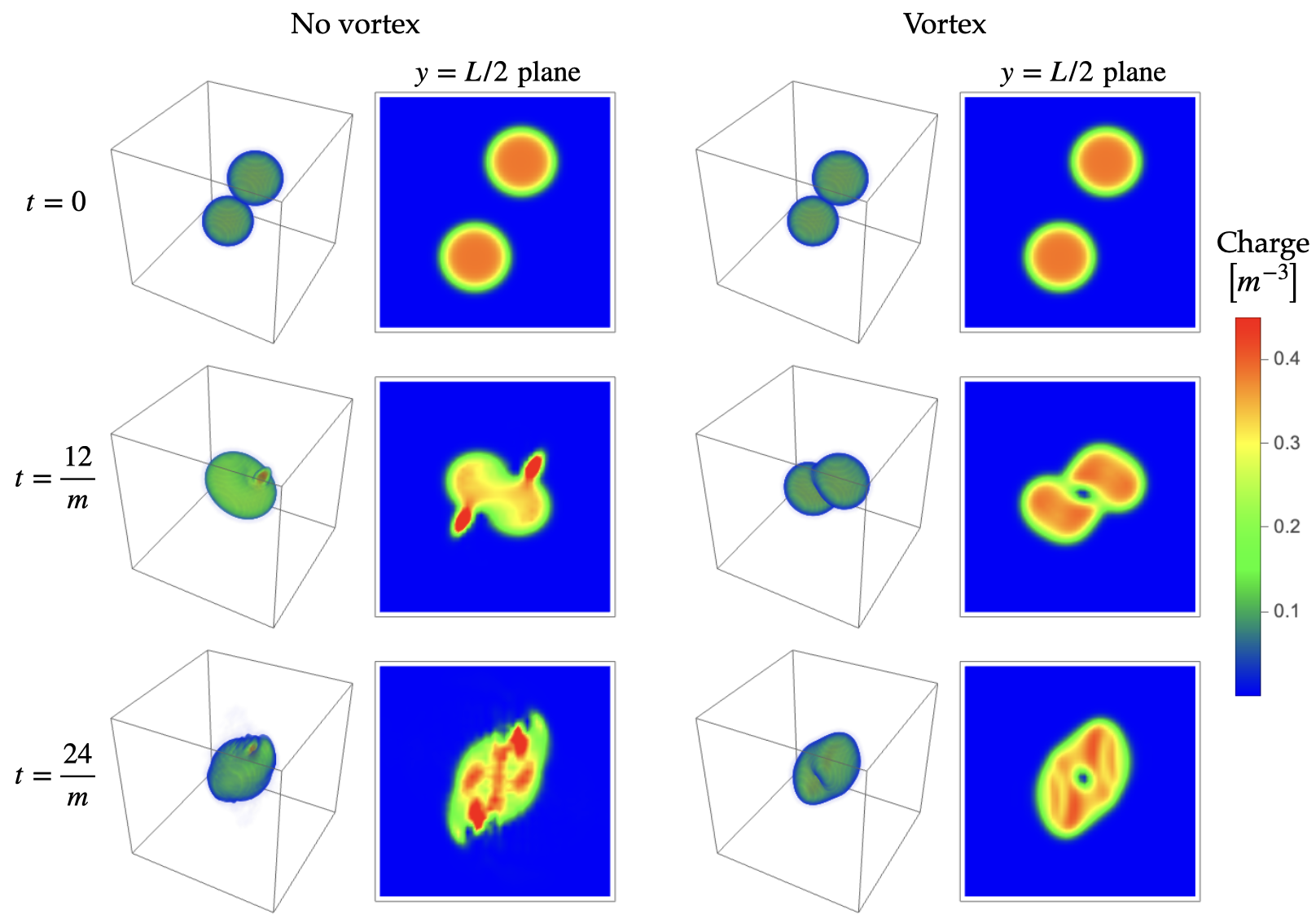}
    \begin{flushleft}
    \caption{
   Snapshots of the charge density from the numerical three-dimensional evolution of the merger dynamics for the no-vortex ({\it left two columns}) and vortex case ({\it right two columns}). Odd columns show the full three-dimensional snapshot, while even columns display the two-dimensional plane clipped at the $Q$-ball centers. In these simulations $\alpha =1$, $\omega = 0.4 m$, $v_{\rm intial}=0.25$ and the impact parameter $b = 12 m^{-1}$.}
    \label{fig:snapshot}
    \end{flushleft}
\end{figure*}

\begin{figure}
    \vs{1.5mm}
    \centering
    \includegraphics[width = 0.9\linewidth]{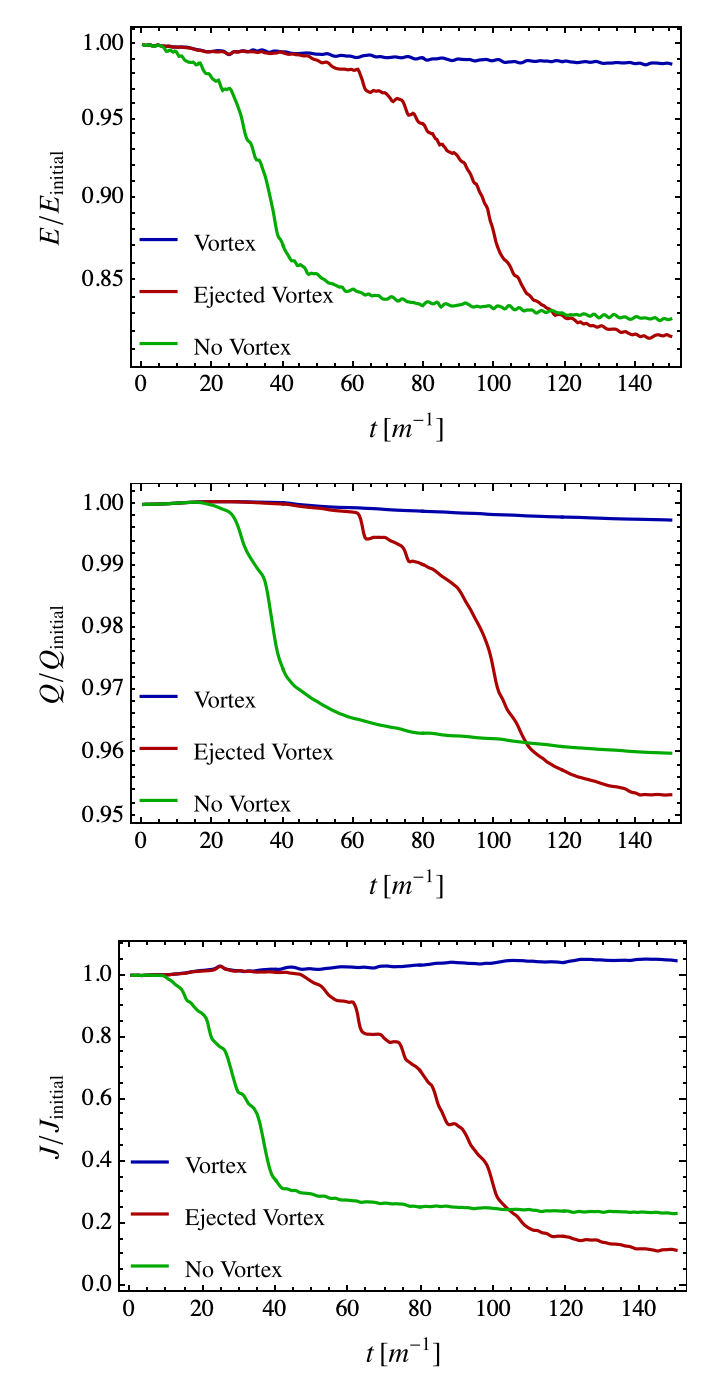}
    \caption{Energy ({\it upper panel}\.), charge ({\it middle panel}\.) and angular momentum ({\it lower panel}\.) as a function of time for the three cases outlined in the text.
    \vs{-3mm}
    }
    \label{fig:energycharge}
\end{figure}

Snapshots of the dynamics for $t = 0,\,12\.m^{-1}$ and $24\.m^{-1}$ can be observed in Fig.~\ref{fig:snapshot} for the vortex and no-vortex cases. A full video of the dynamics can be found at~\href{https://www.youtube.com/watch?v=t29WUvZM-io}{this URL}. The resulting dynamics are summarised by the energy, charge, and spin of the configurations as a function of time in the three cases as shown in Fig.~\ref{fig:energycharge}:
\vs{-1.8mm}
\begin{itemize}

    \item {\it no-vortex regime} (green line): the bubbles merge, radiating a large fraction of their initial energy.\vs{-1.8mm}
    
    \item {\it ejected-vortex regime} (red line): the initial conditions are close to the threshold of vortex formation. The merged soliton exhibits vorticity for a finite amount of time. However, eventually, it ejects the vortex and relaxes to a vortex-free configuration.\vs{-1.8mm}
    
    \item {\it vortex regime} (blue line): a stable vortex, lasting on the timescales of the simulations, is formed within the resulting soliton.\vs{-1.8mm}

\end{itemize}

In the third case, in sharp contrast to the first one, we observe that little to no emission takes place throughout the merger. That is, the would-be emission energy is invested internally into vortex formation. This is also compatible with the previous reasoning that a saturated $Q$-ball with vortex has no available soft-quanta for relaxation~\footnote{This provides further support of the claim of Ref.~\cite{Dvali:2021ofp} that $Q$-ball-type saturons with vorticity cannot deplete their constituents due to the topological quantized nature of the vortex' winding. This is in stark contrast to their non-rotating counterparts.}.

The second case of vortex ejection is intermediate between the two other regimes. The initial conditions are in close vicinity to the threshold of vortex formation. Thereby a finite time interval exists in which a vortex persists in the final bubble. During this phase, no emission takes place, effectively mimicking the vortex case. Correspondingly, the energy evolution of this temporary scenario tracks that of the vortex case. Eventually, the instability of the configuration leads to a rapid emission of the vortex and to the relaxation of the $Q$-ball. A sizable additional energy is emitted in the ejected-vortex regime as compared to the no-vortex case due to the extra misalignment of modes subject to $\Delta \theta \neq 0$.

As can be seen from the upper panel of Fig.~\ref{fig:energycharge}, the behavior of the energy is typical for the relaxation of unstable configurations due to the exponential growth of instability modes. The lost charge in the merger, displayed in the middle panel of Fig.~\ref{fig:energycharge}, is correlated with the loss of energy. 

Finally, the lower panel of Fig.~\ref{fig:energycharge} shows the evolution of the configuration's angular momentum $J$ as a function of time. In the case where the system is away from the vortex formation, a significant portion of the angular momentum is emitted at the merger, as can be seen from the green line. In contrast, in the vortex-formation case, the angular momentum of the configuration slightly increases by $\Ocal( 1 )$ per cent. This, combined with the energy and charge trajectory, is a clear indication of proximity to the classical stationary vortex bubble. In the scenario close to the threshold of vortex formation (red line), the merged saturon behaves analogously to the stable vortex case for a prolonged period of time. Eventually, due to instability, the vortex is ejected from the configuration resulting in an almost complete drop in the final angular momentum. This can be understood from the fact that the vortex was responsible for most of the spin.

What about the case $n > 0$ when a vortex is already present in the initial saturons? Without loss of generality, we focus here on the $2 + 1$ dimensional case. If both incoming saturons possess vorticity with winding $n = 1$, two scenarios are possible in accordance with the previously discussed cases. 

For $0\neq\Delta \theta \ll 1$, a vortex gets localized in the interior of the intermediate saturon, which gets subsequently ejected. This results in a delayed blast of energy.
With increasing $\Delta \theta$ the ejection time shortens, but it remains macroscopic even for macroscopic values of $\Delta\theta$ (smaller than $0.1\,\pi$).

For $\Delta \theta \sim 1$, the vortices are ejected around the time of the merger. Regardless of the initial difference in phases, the remnant has a small angular momentum as compared to the initial configuration (below $20\%$).  

Analogous dynamics are observed in the case of a merger between $n = 0$ and $n = 1$ vortex saturons. Specifically, the vortex is ejected at the merger point, and the final configuration is a pulsating, vortex-free soliton with small angular momentum. In this case, however, independently of the initial $\Delta \theta$, no metastable vortex is observed in the intermediate configuration.

The numerical simulations of these scenarios are shown \href{https://youtu.be/zorkQSYCliU}{HERE}.

\paragraph{\textbf{Consequences for Black Holes.}} So far we have investigated collisions of two saturon bubbles. Three important results emerge. The first one is the very fact of a merger. Secondly, it is possible to form a vortex, even if no vortices were initially present. Thirdly, the vortex formation significantly affects the emission of energy, leading to suppression by several hundred percent compared to the non-vortex case.

In addition, we observe some transient regimes. When close to the vortex-formation threshold, the merged bubble sustains vorticity during an extended period of time. Eventually, the vortex is ejected and a significant fraction of its mass is emitted in radiation. 
 
Two notable features arise in this case: 1) The total emitted radiation is larger when compared to the case of mergers far away from the vortex-formation threshold and 2) the resulting bubble carries close-to-zero angular momentum. The reason for 2) is that the vortex is responsible for the spin of the configuration as can be understood from Eqs.~\eqref{eq:bubbleaxial} and~\eqref{eq:Jt0phi}. In other words, vorticity provides a jump in the angular momentum of the bubble, leading to a slowly-rotating final soliton.

The relevant question is: what conclusions can be drawn for black hole mergers? We must take into account that the spectrum of classical black holes is different from their non-gravitational pendants considered here. Black holes possess a continuous spectrum of axially symmetric configurations of arbitrary angular momentum, while the saturon $Q$-balls in our model do not. 
\newpage

Another question is the rigidity of the connection between the spin and vorticity in the two cases. For the studied $Q$-ball saturons, the spin implies vorticity. This need not necessarily be true for all black holes, especially the black holes of low spin. Rather, the correspondence works very well in the limit of maximal entropy~\eqref{eq:maximalspinsat}. 

From this point of view, we expect that a high probability of vorticity exists for black holes of high spin. Of course, this does not exclude vortex formation even in black holes of a very low spin, as the total winding number can be zero. However, such configurations are likely to be above ground state and thus unstable, but sufficiently long-lived for observational interest.

From the above, we draw the following lesson. The universality of saturation suggests that it is important to study features of vorticity in black hole mergers, taking the features exhibited by saturons as qualitative guidelines.

Treating the saturons in non-gravitational systems as theoretical laboratories for black holes, we are led to the expectation that vorticity could cause similar imprints in black hole mergers. Specifically, if the black hole binary system approaches the vortex-formation threshold at its merger point, we expect a potentially significant delay of the radiation signal. Moreover, the final burst can be expected to be highly energetic{\,---\,}enhanced compared to the case with no vorticity at all{\,---\,}as it carries away most of the angular momentum of the configuration i.e., $J_{\rm radiation} \sim \Ocal( S )$. Therefore, this would also result in a black hole with low spin.

Black hole vortices can be expected for high spins~\cite{Dvali:2021ofp}. These are not uncommon for stellar black holes but predominantly occur in the supermassive range~\cite{Reynolds:2013rva} and not much in the observed black hole mergers of order $\Ocal( 1\text{\,--\,}100 )\.\Msun$~\cite{LIGOScientific:2021djp}. However, a few black holes with a final spin around or less than $10\%$ away from extremality have been observed (such as GW200308). Forthcoming observational runs of LIGO/Virgo/Karga as well as future observatories can be expected to find further mergers with high final spins. These will also probe part of the subsolar mass range in which only primordial black holes~\cite{Hawking:1971ei, Carr:1974nx} could have been formed (see Refs.~\cite{Escriva:2022duf, Carr:2023tpt} for recent reviews). If formed during an epoch of matter domination (cf.~Ref.~\cite{Harada:2016mhb}) or from quark confinement~\cite{Dvali:2021byy}, their angular momenta can be near extremal, particularly in the subsolar mass range. If primordial black holes constitute a substantial fraction of the dark matter, the mentioned mechanisms could yield a large amount of black holes at or near the vortex-formation threshold.

Either from stellar or primordial black holes, the observation of high spins can be expected in the near future. Some of these black holes might carry or generate vorticity, which, as we have argued in this work, would leave observable signatures in forthcoming gravitational-wave searches, unveiling the underlying quantum nature of black holes.

\paragraph*{{Acknowledgements. }}

This work was supported in part by the Humboldt Foundation under Humboldt Professorship Award, by the European Research Council Gravities Horizon Grant AO number: 850 173-6, by the Deutsche Forschungsgemeinschaft (DFG, German Research Foundation) under Germany's Excellence Strategy - EXC-2111 - 390814868, and Germany's Excellence Strategy under Excellence Cluster Origins - EXC 2094 – 390783311.

\noindent\textbf{Disclaimer:}
Funded by the European Union. Views and opinions expressed are however those of the authors only and do not necessarily reflect those of the European Union or European Research Council. Neither the European Union nor the granting authority can be held responsible for them.
\vs{-4.5mm}

\setlength{\bibsep}{4pt}
\bibliography{citations}

\end{document}